\documentclass[acmlarge]{acmart}
\AtBeginDocument{%
  \providecommand\BibTeX{{%
    \normalfont B\kern-0.5em{\scshape i\kern-0.25em b}\kern-0.8em\TeX}}}

 \setcopyright{acmcopyright}
 \copyrightyear{2023}
 \acmYear{2023}
 \acmDOI{}

\copyrightyear{2023}
\acmYear{2023}
\setcopyright{rightsretained}
\acmConference[CHI'23 1st Workshop on Behavioural Design in Video Games]{CHI Conference on Human Factors in Computing Systems}{April 23-28, 2023}{Hamburg, Germany} \acmBooktitle{CHI'23 1st Workshop on Behavioural Design in Video Games: Ethical, Legal, and Health Impact on Players, April 23-28, 2023, Hamburg, Germany} \acmPrice{} \acmDOI{} \acmISBN{}

\begin{document}

\title{Time-Based Addiction}


\author{Ziwei Gao}
\email{annannika.biu@gmail.com}
\affiliation{%
  \institution{University College London}
  \city{London}
  \country{UK}
}


\renewcommand{\shortauthors}{Gao}

\begin{abstract}
This paper introduces time-based addiction, which refers to excessive engagement in an activity that results in negative outcomes due to the misallocation of time. This type of addiction is often seen in media-related activities such as video games, social media, and television watching. Behavioural design in video games plays a significant role in enabling time-based addiction. Games are designed to be engaging and enjoyable, with features such as rewards, leveling up, and social competition, which is all intended to keep players coming back for more. This article reviews the behavioural design used in video games, and media more broadly, to increase the addictive nature of these experiences. By doing so the article aims to recognise time-based addiction as a problem that in large part stems from irresponsible design practices. 
\end{abstract}




\maketitle

\section{Introduction}

Addiction is a complex phenomenon that has long been studied in various forms, including substance-related addictions and behavioural addictions. This paper introduces time-based addiction, which refers to excessive engagement in an activity that results in negative outcomes due to the misallocation of time. This type of addiction is often seen in media-related activities such as video games, social media, and television watching. With the rise of new media and mediums, it is becoming increasingly important to understand the nature and effects of time-based addiction. In this article, we aim to explore the concept of time-based addiction, its causes, and its potential impacts on individuals. It will primarily explore ways in which UX and UI design can promote this class of addiction. 

Behavioural design in video games plays a significant role in enabling time-based addiction. Games are designed to be engaging and enjoyable, with features such as rewards, leveling up, and social competition, which is all intended to keep players coming back for more. These features are designed to create a sense of progress and achievement, which can be highly motivating for players. However, when these features are combined with other addictive elements, such as endless gameplay or the option to pay for in-game items, they can lead to excessive engagement, or worse, a time-based addiction.

Algorithms are also increasingly used to keep players engaged in a game over longer periods. The use of algorithms to optimise the game's difficulty, keep players engaged, and encourage them to spend more time playing, only exacerbates the problem. The result is a highly engaging and addictive experience that can cause players to lose track of time and neglect other important aspects of their lives - or in other words, to misallocate their time. Thus, behavioural design in video games is a crucial factor in the development of time-based addiction.

In this article, we will introduce and define the concept of time-based addiction. We will then move on to discussing media consumption as a very common type of time-based addiction. Video games and the behavioural design involved in making them will serve as the main case study for a very common type of time-based addiction. Time-based addiction will also be discussed beyond the realm of media consumption in order to draw parallels and broaden our understanding of the concept. Finally, \textit{time loss aversion} will be discussed as a potential solution to solving video game addiction with the use of behavioural design.

\section{Defining Time-Based Addiction}

Time-based addiction refers to a pattern of excessive engagement in an activity, such as social media use, video gaming, or streaming, that results in negative outcomes due to the misallocation of time. This type of addiction is different from more biological substance-related addictions, where the focus is on the use of a particular substance, such as drugs or alcohol. Instead, the focus of time-based addiction is on the excessive engagement in an activity that leads to an individual losing track of time and neglecting other important aspects of their life. For example, someone who spends hours playing video games each day may miss out on sleep, exercise, or other leisure activities. They may also neglect responsibilities such as work, school, or relationships, leading to negative life outcomes. Additionally, time-based addiction may end up causing physical problems, as well as mental health issues and social isolation. Thus, time-based addiction is a serious issue that needs to be recognised and addressed. Finally, time-based addiction can result in \textit{virtual spillover} - the adoption of preferences and behavior while engaging with online media which then impact people's preferences and behavior in other aspects of their lives \cite{franklin2022virtual}. An example of this would be impulsiveness or a preference for instant gratification.

From this, we can conclude that time-based addiction is a behavioural addiction rather than a substance addiction \cite{james2017need}. Thus time-based addiction is similar to other behavioural addictions such as gambling, with the crucial difference being that gambling also involves potential monetary rewards, and the negatives associated with it are not only the misallocation of time but also the misallocation of money. Like with other behavioural addictions, time-based addiction leads to an excessive engagement in an activity that can lead to a sense of escapism and pleasure. This can create a feedback loop, as individuals engage in the activity more and more to avoid the unpleasant feelings of reality, that is getting worse and worse due to the negatives of over-engaging in the addictive behaviour. We argue that time-based addiction is getting worse due to the availability and accessibility of digital media, which has dramatically increased with the rise of consumer electronics and mobile technology \cite{twenge2019trends}. The push to extended, virtual, and augmented reality may only make things worse \cite{franklin2022virtual}. 

As younger people are the early adopters of novel technologies, time-based addiction may have a stronger impact on young people \citep{ruggeri2018new}), who are often more likely to have longer screen times \cite{riehm2019associations, perrin2015social}. Younger people also have worse inhibition and self-regulation skills \cite{ordaz2013longitudinal}. Further, due to the fact that young people are establishing themselves and who they want to be, the misallocation of time will result in uniquely negative outcomes and set bad habit formation that can plague them in the future. 

\section{Video Games: A case study for time-based addiction}

Video game addiction has been recognised as a growing concern in recent years, with numerous studies pointing to its negative impacts on mental (e.g., stress, depression, anxiety) \cite{loton2016video, weinstein2010computer} and physical health (e.g., musculoskeletal issues, vision problems, and obesity) \cite{naser2016detecting, stockdale2018video}. Younger people are at a higher risk as they play longer and are more sociable \cite{entwistle2020video}. We argue that video game addiction is a prime example of time-base addiction.

Research has argued that long hours on video games can be exacerbated by "ineffective time management skills, or as a symptomatic response to other underlying problems that they are escaping from, rather than any inherent addictive properties of the actual games \cite{wood2008problems}." Others, on the other hand, have pointed towards the gamified nature of video games, with their emphasis on levels, points, and rewards, as a key contributor to their addictive potential \cite{harrigan2010addictive}.

Gamification refers to the use of game elements and mechanics in non-game contexts (e.g., education or health) to increase engagement and motivation \cite{rodrigues2016gamification, nah2014gamification}. The use of rewards, competition, and progress are used as motivators for players \cite{zichermann2011gamification}. However, the integration of game-like elements has raised concerns about the potential for addictive behaviour \cite{bassanelli2022gamification}. Video games are designed to be engaging and enjoyable, which can lead to addiction \cite{dey2016gamification}. Understanding the relationship between gamification and addiction is crucial for using gamification responsibly and ethically.

Points and rewards are central aspects of the behavioural design of video games, playing a crucial role in their addictive nature \cite{schell2008art}. Rewards are a key aspect of behavioural design in video games, as they provide players with a tangible sense of progress and motivation to continue playing. Rewards can take many forms, such as points, currency, experience points, unlockable items or abilities, and more. The type and frequency of rewards can be used to control the pacing of the game and the player's overall motivation. The reason rewards influence behaviour is that they serve immediate feedback for the player's actions and provide a sense of accomplishment and progress \cite{felicia2012motivation}. These tangible rewards encourage players to continue playing, as they strive to accumulate more points and achieve higher levels.

The constant pursuit of rewards creates a cycle of reinforcement, where players feel more motivated to keep playing in order to receive more rewards \cite{chumbley2006affect}. Additionally, the variable-ratio schedule of rewards in many videogames provides intermittent reinforcement, which has been shown to be more effective in promoting continued behaviour compared to consistent reinforcement \cite{king2011role}. The design creates a feeling of unpredictability and excitement, further driving players to continue playing in the hope of receiving more rewards \cite{king2010video}.

Social behavioural design can increase video game addiction. Players with more problematic video game use tend to prefer massively multiplayer online role-playing games (MMORPGs) \cite{mentzoni2011problematic}. MMORPGs are a type of video game that allows large numbers of players to participate in a virtual world simultaneously and engage in role-playing activities, such as questing and character progression \cite{bartle2004designing}. These games often feature complex systems and mechanics that require strategic planning and cooperation between players, leading to a rich and immersive gaming experience. The additional social media-like behavioural design of MMORPGs makes them more addictive for players. 

Behavioural design in video games not only hooks players into playing for longer periods of time, but also encourages them to prioritise the game over other important aspects of their lives, such as eating, sleeping, and socialising \cite{williams2006groups, peracchia2018exposure, wolfe2014single, arvaniti2011salty, ballard2009correlates}. In extreme cases, this addiction has even resulted in death, as some individuals have played for days on end without taking breaks for basic necessities like food and water \cite{kuperczko2022sudden}.

\section{Media Addiction as a Type of Time-Based Addiction}

Arguably, video game addiction sits within a broader category of time-based addiction - addiction to media. This involves TV watching, social media, streaming, and online content.

TV addiction refers to the excessive consumption of television to the point where it negatively impacts a person's daily life and relationships. The behavioural design of television, such as its repetitive and passive nature, can contribute to TV addiction by providing instant gratification and a sense of escapism. Specifically, repetitiveness refers to repetitive patterns, such as the same cast of characters and the same types of stories, which can make it difficult for viewers to stop watching\citep{kubey2002television}. Passiveness means that viewers can simply sit back and watch the content without actively participating. This passive consumption can make it easy for viewers to become absorbed in the content and lose track of time\citep{condry2017psychology}. Escapism refers to Television providing a form of escapism that allows viewers to escape from their daily lives and immerse themselves in a different world\citep{shanahan1999television}. Television provides instant gratification and a sense of relaxation, which can make it difficult for viewers to stop watching once they have started\citep{condry2017psychology}. 

Online streaming and content platforms, such as Netflix and YouTube can lead to addictive behaviour \cite{balakrishnan2017social}. Research has identified several behavioural design mechanisms that can contribute to online streaming addiction, including unlimited choice, auto-play, and social comparison. Unlimited choice is the seemingly endless options for video content  that are available on streaming and content platforms \cite{chaudhary2022you}. This is combined with the immediate gratification included in streaming platforms allowing users to access content instantly, without any wait time or delay. Online streaming platforms constantly add new content, keeping the platform fresh and exciting for users. This constant influx of new material can create a sense of novelty and can contribute to excessive consumption. Auto-play is a feature that automatically plays the next episode or video in a series, reducing the effort required for the user to continue watching \cite{zundel2019serial}. Social comparison is another factor that contributes to online streaming engagement as it allows users to see what their friends are watching and to share their own viewing habits \cite{zimmer2021closer}.

Social media addiction refers to the excessive use of social media to the point where it negatively impacts a person's daily life and relationships. There are several behavioural design elements that can contribute to social media addiction. First, recommender systems, social media platforms use algorithms to recommend content that is likely to be of interest to a user, which can keep them engaged for longer periods of time\citep{resnick1997recommender}. Recommender systems do not only change people's behaviours on these platforms but over time also change their preferences for what they are interested in \cite{ashton2022problem, franklin2022recognising, ashton2022solutions}. Second, infinite scroll, the continuous scrolling feature on social media platforms makes it difficult for users to stop using the platform, as there is always more content to view \citep{cara2019dark}. Third, constant notifications, social media platforms are designed to send notifications to users, alerting them to new updates and messages. This constant stream of notifications can make it difficult for users to resist checking their social media accounts\citep{kushlev2016silence}. Fourth, instant gratification, social media provides instant gratification through likes, comments, and other forms of social validation, which can create a positive feedback loop and increase usage\citep{du2019predictors}. Finally, Pressure to be Connected: Social media often creates pressure to be constantly connected and up-to-date, which can lead to excessive usage and addiction\citep{vorderer2016permanently}. These behavioural design elements can contribute to social media addiction by creating a cycle of engagement that can be difficult for users to break.

It appears that overall there are similarities and uniqueness to addictive behavioural design in different types of media. The similarities are the escapism, endless choice, and updating of these different media. Social media and streaming services are unique in that they tailor themselves to a particular user whilst TV does not.  TV has a linear design, with a fixed schedule of programming that users watch in real-time. Online streaming services, on the other hand, allow users to watch content on demand, while social media platforms offer a combination of real-time and on-demand content.  TV is primarily a one-way communication medium, with users passively receiving content. Online streaming and social media, on the other hand, are more interactive and allow for two-way communication. The design of social media platforms is centered around user-generated content. TV and online streaming platforms, on the other hand, are centered around professionally produced content. Social media platforms have a strong focus on social interaction, allowing users to connect with others, share content, and engage in discussion. TV and online streaming in comparison have less of a focus on social interaction.

Video games as a type of media addiction share plenty of behaviour design features with other types of media. The way in which it stands out the most is it is a highly active form of media behaviour, and has a higher amount of in-game rewards and progression. To give an example, video games are the only media where one can formally complete the game or "win". This will make the behavioural design of games unique compared to other media. 

\section{Time-based addiction beyond media}

We have argued that media consumption is a form of time-based addiction. However, there are many other behavioural addictions that can be classified as time-based addictions in that they primarily stem from a misallocation of time. This includes work addiction \cite{sussman2012workaholism}, gambling addiction \cite{blanco2001pathological}, and exercise addiction \cite{landolfi2013exercise}. These types of addictions differ from media-based addictions in terms of their specific object of addiction, but they share many similarities in terms of the behavioural triggers that may be involved. 

Like media time-based addictions, non-media time-based addictions also often provide an escape from reality and offer the potential for rewards, such as the sense of accomplishment from a hard day's work, the thrill of winning at gambling, or the physical and emotional benefits of exercise. However, there are also important differences between these types of addictions. For example, work addiction provides financial rewards, which can have significant practical implications, whereas media-based addictions tend to be more focused on virtual rewards, such as points and level-ups.

Crucially, an important difference between media addictions and non-media time-based addictions is the ease of access. The widespread availability of digital devices and the internet has made media-based addictions more accessible and convenient than ever before. Non-media addiction, such as work addiction, may require more effort to maintain \cite{maslach2016burnout}. Understanding these similarities and differences is crucial in addressing and treating time-based addictions in all their forms.

\section{Time Loss Aversion as a Solution to Time-Based Addiction}

“Time Loss Aversion” is people’s tendency to fear previous misallocations of time, which then motivates people to change their future behaviour \cite{gao2023aversion}. Loss aversion has previously been achieved through framing outcomes as losses \cite{tversky1989rational}. Previous research has induced loss aversion by framing monetary outcomes as losses rather than gains \cite{tversky1991loss}. Research has found that when monetary outcomes are framed as losses rather than gains people behave in a more risk-seeking way \cite{tversky1992advances}. Other studies have also induced a sense of loss for non-monetary outcomes. For example, when a public health policy outcome is framed in the number of lives lost rather than saved, it results in loss aversion and increased risk-seeking \cite{kahneman1984choices}. From reviewing the previous literature, it becomes clear that framing time as a loss has been underexplored and underutilised as a potential behaviour change method.

Time loss aversion has the potential of addressing time-based addiction. One study used an approach rooted in time loss aversion where participants were told how their daily social media use aggregated to the level of a week, month of year \cite{gao2023social}. This went beyond usual approaches that would show people their average "screen time" on different apps. Participants' intention to use social media in the future significantly decreased and their intention to engage with non-digital, non-addictive behaviours increased. The decrease in intended future social media use dropped by a whole hour on average. Heavy social media users - defined as users that would use 4 or more social media apps regularly - would further benefit from being told how they haven't been spending their time. For example, in addition to being told the aggregation of their daily use, they would be told how that time could have been spent differently (e.g., seeing friends or family). This is the other side of time loss aversion - people presumably weren't given any new information about how they can spend their time, but rather the fear of what they haven't been doing influenced their behaviour.

\section{Time-based addiction and FOMO}

It is important to note however that time loss aversion might also be used to increase people's engagement in addictive media such as video games. The "fear of missing out" (FOMO) can be seen as a type of time loss aversion. Studies have shown that people who were not actively engaging with social media sometimes felt anxiety if they knew that others were currently spending their time on social media without them in a way that felt like a rewarding social experience. This was especially true if it was people within their social network. More broadly, studies find that increased FOMO relates to increased mobile phone use across the board \cite{wolniewicz2018problematic, coskun2019investigation}. Thus, time loss aversion is a double-edged sword in that it can promote media addiction if used in a specific way.

FOMO is also a problem for video games. One study found that players of MMORPGs were especially prone to this \cite{duman2021impact}. MMORPGs have a behavioural design that borrows from both social media and games.  Indeed, the study found that players social identities mediated their FOMO when not engaging in the game. Another study founds that mobile game marketing and communication practitioners were aware of FOMO and video games \cite{perreault2020mobile}. The messages and strategies employed by these practitioners when creating promotional materials seek to induce the feeling of FOMO. Certain players are at a higher risk of developing time-based addictions to games due to FOMO. One study on a large sample of Chinese university students found that impulsivity mediated the relationship between excessive gaming and FOMO, and that there was a larger effect size of this mediation for male students \cite{li2021mediating}. 

\section{Conclusion}

The goal of this paper was to recognise and define time-based addiction as a problem that is still in its early stages. We need more research and education to better understand its causes and consequences. It is evident from the analysis in this article that behavioural design in video games has increased the addictive potential of video games. Further, other forms of media have borrowed from behavioural design in video games in the form of gamification. This on its own points towards an intuition shared by researchers and practitioners - that video game design if used irresponsibly may be a large source of time-based addiction. The negative outcomes of time-based addictions are negative on both and individual and societal level, thus more research into its causes and prevention are needed.

\bibliographystyle{ACM-Reference-Format}
\bibliography{sample-base}


\end{document}